\documentclass[
superscriptaddress,
reprint,
nofootinbib,
nobibnotes,
eprint,
amsmath,
amssymb,
aps,
prb,
citeautoscript,
nobalancelastpage,
floats,
final,
eqsecnum
]{revtex4-2}

\usepackage[utf8]{inputenc}

\usepackage{amsmath}
\usepackage{amssymb}
\usepackage{amsfonts}

\usepackage{latexsym}
\usepackage{slashed}
\usepackage[svgnames]{xcolor}
\usepackage{soul}
\usepackage{braket}
\usepackage{MnSymbol}
\usepackage{wasysym}
\usepackage{mathrsfs}  
\usepackage{physics}
\usepackage{enumitem}      
\usepackage[title]{appendix}
\usepackage{tikz}

\usepackage{graphics}
\graphicspath{ {graphs/} }

\usepackage{microtype}
\usepackage{multirow}

\usepackage{hyperref}
\usepackage{natbib}

\hypersetup{
	pdfencoding=auto,
	psdextra,
	colorlinks=true,
	linkcolor=Blue,
	citecolor=DarkRed,
	urlcolor=Blue
}

\usepackage{mathtools}

\usepackage{easyReview}

\begin{document}
	
	\title{\Large{{\sc Parafermionic formulation of a  one-dimensional generalized $N$-state clock model }}}
	
	\author{Gustavo M. Yoshitome}
	\email{gustavoyoshitome@gmail.com }
	\affiliation{Departamento de Física, Universidade Estadual de Londrina, Londrina, PR, Brasil}
    \affiliation{Instituto Federal do Paraná,
Campus Astorga, Astorga, PR, Brasil}
	

\begin{abstract}	

In this work, we investigate a parafermionic formulation of a one-dimensional generalized $N$-state clock model, a modification of the conventional clock model controlled by an integer parameter $a$. We determine how these parameters enter the construction of a local parafermionic description through the Fradkin-Kadanoff transformation. A suitable change of basis, together with a factorization of the local Hilbert space, allows us to identify the sector supporting $\mathbb{Z}_{N_a}$ parafermions, where $N_a$ is determined by $N$ and $a$. At an exactly solvable point, we identify parafermion zero modes localized at the ends of the open chain and show that their order is precisely $N_a$. We further interpret this dependence by realizing the generalized clock model at the boundary of a generalized toric code. Finally, we examine the Kramers-Wannier duality in the parafermionic formulation and show how it acts as a translation of parafermion bilinears relating the original and dual chains. 

\end{abstract}

\maketitle

\tableofcontents

\newpage

\section{Introduction}

Recently, a special class of symmetries known as modulated symmetries has attracted attention because these symmetries appear in many $d$-dimensional lattice models \cite{PhysRevB.105.045128,Delfino:2022ndx,PhysRevB.106.045145,Delfino:2023anb,Chen:2025hzm,Yoshitome:2025klz,vvhq-wnll}, and provide insight into properties such as the structure of the ground state subspace and the mobility of excitations, particularly in two-dimensional symmetry-enriched topologically ordered phases that support anyons \cite{Pace:2025hpb,Bulmash:2025zua,Ning:2026noc,Pace:2026brj,Yoshitome:2026qzp}. The striking feature of these symmetries is the nontrivial action of spatial transformations on the internal symmetry group. The toric code, a paradigmatic example of topological order, can be generalized to exhibit modulated symmetries \cite{Watanabe:2021wwt}.

In one dimension, a particularly interesting model exhibiting modulated symmetry is the generalized $N$-state clock model \cite{PhysRevB.107.195139}. This model is a generalization of the $N$-state clock model controlled by an integer parameter $a$. Its symmetry leads to a ground state degeneracy that depends on the lattice size. For particular values of $N$ and $a$, the model can undergo spontaneous symmetry breaking without ground state degeneracy. Moreover, the model admits a non-invertible Kramers-Wannier duality, analogous to that of the transverse-field Ising model \cite{Seo:2024its,Pace:2024tgk}.

It is known that the transverse-field Ising model can be mapped to a chain of Majorana fermions, known as the Kitaev chain, through a Jordan-Wigner transformation \cite{Kitaev:2001,Kitaev:2009war,Alicea_2012}. For the $N$-state clock model, an analogous transformation can be performed in terms of parafermions \cite{FRADKIN19801,Fendley:2012vv,Mong:2014ova,li2015criticality,Yao:2020dqx}, which are particles that generalize Majorana fermions and emerge as defects of fractionalized systems \cite{clarke2013exotic,PhysRevB.86.195126,PhysRevX.2.041002,PhysRevB.95.075111,Dua:2018vja,bollmann2026phases,Yoshitome:2025cit,zheltonozhskii2026classification}, revealing properties such as parafermion zero modes in open chains. Although this transformation can be applied to any Hamiltonian expressed in terms of $\mathbb{Z}_N$ clock variables, it does not always yield a local Hamiltonian in the parafermion variables.

In this work, we investigate whether a parafermionic formulation of the generalized $N$-state clock model is possible using the Fradkin-Kadanoff transformation. We verify that a naive transformation of the clock variables does not result in a local Hamiltonian, but introducing a suitable change of basis, this transformation yields a local Hamiltonian in the parafermion variables.

A distinctive feature of the parafermionic formulation of this model is that locality is achieved by introducing $\mathbb{Z}_{N_a}$ parafermions instead of $\mathbb{Z}_N$ parafermions, where $N_a$ is the largest divisor of $N$ that is coprime to $a$. When $N$ and $a$ are coprime, then $N=N_a$, otherwise, the parafermions acquire a reduced order. In this case, only a sector of the full Hilbert space is expressed in terms of parafermions. We argue that this is expected when the one-dimensional model is interpreted as the boundary of the two-dimensional generalized toric code.

We then discuss how the Kramers-Wannier duality of the generalized $N$-state clock model manifests in the parafermionic Hamiltonian. In the transverse-field Ising model, this duality corresponds to a translation by one Majorana operator, which can be understood as a half-lattice translation in the spin description. Here, this interpretation must be adapted by introducing a dual parafermionic chain obtained by applying the Kramers-Wannier transformation to the generalized $N$-state clock model.

This work is organized as follows. In Sec. \ref{sec2}, we review some relevant aspects of the generalized $N$-state clock model, including its ground states under open and periodic boundary conditions. In Sec. \ref{IV}, we construct a parafermionic formulation of the model when the integers $a$ and $N$ are coprime. In Sec. \ref{secV}, we show how this parafermionization can be extended to general values of $N$ and $a$. In Sec. \ref{secVI}, we establish the correspondence between the parafermions and the topological order of a generalized toric code. Finally, in Sec. \ref{secVII}, we discuss how the Kramers-Wannier duality of the generalized $N$-state clock model manifests in the parafermionic formulation.



\section{Aspects of the model\label{sec2}}

In this section, we introduce the generalized $N$-state clock model, a generalization of the Ising model, defined on a one-dimensional lattice of $L$ spins, and some of its properties, such as symmetry operator and ground state subspace.

The Hamiltonian of the generalized $N$-state clock model is a sum of local operators expressed in terms of $\mathbb{Z}_N$ clock variables $Z_j$ and $X_j$, satisfying
\begin{equation}
    Z_jX_{j^\prime} = \omega^{\delta_{j,j^\prime}} X_{j^\prime}Z_j, \qquad Z_j^N = X_j^N = 1,
\end{equation}
where $\omega = e^{\frac{2\pi i}{N}}$. The full Hilbert space is $\mathcal{H} = (\mathbb{C}^N)^{\otimes L}$, and under periodic boundary conditions (PBC), the Hamiltonian is
\begin{equation}
\label{3.1}
    H = - \sum_{i=1}^{L} Z_i^{-a} Z_{i+1} -g\sum_{i=1}^L X_i+\mathrm{h.c.},
\end{equation}
where $a$ is an integer parameter. For $a=1$, the model reduces to the usual clock model. Under open boundary conditions (OBC), the upper limit of the first sum is $L-1$.

Under PBC, the model has a global symmetry generated by $U=W^n$, where
\begin{equation}
\label{3.2}
    W = \prod_{i=1}^{L} X_i^{a^{i-1}}, \qquad n = \frac{N}{\gcd(a^L-1,N)}.
\end{equation}
The exponent $n$ arises from the boundary conditions, since $W$ generally does not commute with the last bond interaction, except when $\gcd(a^L-1,N) = N$.

This symmetry is known as an exponential symmetry and is spatially modulated; that is, it transforms nontrivially under translation. Note that under a translation, $Z_i \longmapsto Z_{i+1}$ and $X_i \longmapsto X_{i+1}$, the global internal symmetry transform as
\begin{equation}
    U \longmapsto U^{a^{-1}}, 
\end{equation}
\footnote{Here, $a^{-1}$ is the inverse of $a$ modulo $\gcd(a^L-1,N)$.}Thus the symmetry group is $G = \mathbb{Z}_{\gcd(a^L-1,N)} \rtimes T$, where $T$ is the spatial group of translations  \cite{Yoshitome:2026qzp}. Models exhibiting this type of symmetry often have properties that depend on microscopic details of the lattice. 

This model exhibits spontaneous symmetry breaking, with a ground state degeneracy given by $\mathrm{GSD} = \gcd(a^L-1,N)$:
\begin{itemize}
    \item[1.] When $a^L-1$ and $N$ are coprime, the ground state is unique.
    \item[2.] When $\gcd(a^L-1,N) = N$, the ground state degeneracy is $N$, as in the usual clock model.
\end{itemize}

One of the distinctive features of this model is that it allows a phase transition into a symmetry-broken phase without ground state degeneracy \cite{PhysRevB.107.195139}.

\subsection*{Ground states for OBC and PBC}
The dependence of the ground state degeneracy on the boundary conditions can be understood explicitly at the exactly solvable point $g=0$. At this point, the Hamiltonian is diagonal in the clock basis, and its ground states are determined by constraints relating neighboring spins. Under OBC, these constraints can be satisfied recursively, whereas PBC impose an additional consistency condition that depends on $N$, $a$, and the chain length $L$.
We construct each ground state explicitly. If
\begin{equation}
\ket{\psi} = \ket{n_1 \,\, n_2 \,\, \cdots \,\, n_{L}}, \qquad Z_l\ket{n_l} = e^{\frac{2\pi i n_l}{N}}\ket{n_l}
\end{equation}
is a ground state, then the ground state condition implies
\begin{equation}
\label{34}
    n_{l+1} = a\,n_l \bmod{N}.
\end{equation}
Under OBC, we can choose $n_1 = 0,1,\ldots,N-1$, and the recurrence relation fixes the remaining values of $n_l$. This yields a total of $N$ ground states. Under PBC, however, an additional constraint relates the first and last spins of the chain:
\begin{equation}
    n_1 = a\,n_L \bmod{N}.
\end{equation}
From Eq. (\ref{34}), we have $n_L = a^{L-1}n_1 \bmod{N}$. Therefore,
\begin{equation}
    n_1(a^L-1) = 0 \bmod{N}.
\end{equation}
The number of solutions to this equation is exactly $\gcd(a^L-1,N)$, which gives the ground state degeneracy under PBC.

It is not obvious whether the Fradkin-Kadanoff transformation, which maps clock operators to parafermion operators \cite{FRADKIN19801,Fendley:2012vv}, can yield a local parafermionic formulation of the Hamiltonian (\ref{3.1}), analogous to the Jordan-Wigner mapping of the transverse-field Ising model to the Kitaev chain \cite{Kitaev:2001}. Below, we show that such a formulation is indeed possible through a suitable change of basis.


\section{Parafermion transformation when $a$ and $N$ are coprime  \label{IV}}

In this section, we focus on the case $\gcd(a,N) = 1$ and show that the model can be written as a local Hamiltonian of parafermions. The general case, without requiring $a$ and $N$ to be coprime, is discussed in Sec. \ref{secV}. We begin by redefining the original clock operators as
\begin{equation}
    \mathcal{Z}_i = Z_i^{a^{-i}}, \qquad \mathcal{X}_i = X_i^{a^i},
    \label{4.1}
\end{equation}
where $a^{-i}$ denotes the multiplicative inverse of $a^i$ modulo $N$. These operators still satisfy the clock algebra,
\begin{equation}
    \mathcal{Z}_j\mathcal{X}_{j^\prime} = \omega^{\delta_{j,j^\prime}} \mathcal{X}_{j^\prime}\mathcal{Z}_j.
    \label{3.3}
\end{equation}
In terms of these new clock operators, the Hamiltonian (\ref{3.1}) has spatially dependent exponents:
\begin{equation}
    H = - \sum_{i=1}^{L-1} (\mathcal{Z}_i^{\dagger} \mathcal{Z}_{i+1})^{a^{i+1}} - \mathcal{Z}_{L}^{-a^{L+1}}\mathcal{Z}_1^a -g\sum_{i=1}^L \mathcal{X}_i^{a^{-i}}+\mathrm{h.c.}
\end{equation}
However, the symmetry operator takes a spatially uniform form,
\begin{equation}
    U = W^n, \qquad W = \prod_{i=1}^{L}\mathcal{X}_i^{a^{-1}}.
    \label{3.4}
\end{equation}

The exponent $n$ accounts for the periodic boundary conditions, as $W$ does not commute with the boundary bond interaction unless $\gcd(a^L-1,N) = N$, whereas $W^n$ does. Now, the Hamiltonian is not invariant under a simple translation, instead, it is invariant by
\begin{equation}
\begin{cases}
    \mathcal{Z}_i \longmapsto \mathcal{Z}_{i+1}^{a}, \quad \quad \mathcal{X}_i \longmapsto\mathcal{X}_{i+1}^{a^{-1}}\\
    \\
    \mathcal{Z}_L \longmapsto\mathcal{Z}_1^{a^{1-L}}, \quad \quad \mathcal{X}_L\longmapsto\mathcal{X}_1^{{a^{L-1}}}.
\end{cases}
\end{equation}
This transformation maps the uniform symmetry $U$ to $U^{a^{-1}}$. Therefore, the symmetry group remains unchaged, $G =\mathbb{Z}_{\gcd(a^L-1,N)} \rtimes T$.

\subsection{Open Boundary Conditions}

Under OBC, the model takes a simpler form because the interaction connecting the last and first sites is absent. The Hamiltonian becomes
\begin{equation}
    H = - \sum_{i=1}^{L-1} (\mathcal{Z}_i^{\dagger} \mathcal{Z}_{i+1})^{a^{i+1}} -g\sum_{i=1}^{L} \mathcal{X}_i^{a^{-i}}+\mathrm{h.c.}
    \label{4.6}
\end{equation}
The operator $W$ in Eq. (\ref{3.4}) is now a symmetry for all coprime values of $a$ and $N$. At $g=0$, the ground state degeneracy is $N$, and $W$ permutes between the $N$ ground states, with $W^N=1$.

The clock model can be parafermionized by expressing the clock variables in terms of parafermion operators through the Fradkin-Kadanoff transformation:
\begin{equation}
    \alpha_{2j-1} = \left(\prod_{k<j} \mathcal{X}_k\right) \mathcal{Z}_j, \qquad j \geq 1,
    \label{4.9}
\end{equation}
\begin{equation}
    \alpha_{2j} = \omega^{\frac{N+1}{2}} \left(\prod_{k<j} \mathcal{X}_k\right) \mathcal{Z}_j\mathcal{X}_j, \qquad j \geq 1.
    \label{4.10}
\end{equation}
These operators satisfy the parafermion algebra
\begin{equation}
    \alpha_j\alpha_{j^\prime} = \omega^{\operatorname{sgn}(j^\prime-j)}\alpha_{j^\prime}\alpha_j.
\end{equation}
In particular, when $N=2$, they reduce to Majorana operators, which anticommute for distinct $j$ and $j^\prime$. In terms of these operators, the Hamiltonian in Eq. (\ref{4.6}) becomes
\begin{eqnarray}
    H = &-& \sum_{i=1}^{L-1} \left(-\omega^{\frac{3}{2}}\alpha_{2i}^\dagger\alpha_{2i+1}\right)^{a^{i+1}}+\\ &-& g \sum_{i=1}^{L}\left(-\omega^{-\frac{1}{2}}\alpha_{2i-1}^\dagger\alpha_{2i}\right)^{a^{-i}} + \mathrm{h.c.}
\end{eqnarray}

In the usual clock model, the $N$-fold ground state degeneracy at $g=0$ is associated with parafermion zero modes at the ends of the chain. The same holds here: at $g=0$, the two end parafermions commute with the Hamiltonian and connect the $N$ degenerate ground states.

\subsection{Periodic Boundary Conditions}

Under periodic boundary conditions, the Hamiltonian takes the form
\begin{eqnarray}
\nonumber
    H = &-& \sum_{i=1}^{L-1} (\mathcal{Z}_i^{\dagger} \mathcal{Z}_{i+1})^{a^{i+1}}\\ &-& \mathcal{Z}_{L}^{-a^{L+1}}\mathcal{Z}_1^a -g\sum_{i=1}^L \mathcal{X}_i^{a^{-i}} + \mathrm{h.c.},
    \label{4.13}
\end{eqnarray}
which we reproduce here for convenience.

Applying the Fradkin-Kadanoff transformation defined in Eqs. (\ref{4.9}) and (\ref{4.10}) yields
\begin{eqnarray}
\label{4.12}
\nonumber
    H = &-& \sum_{i=1}^{L-1} \left(-\omega^{\frac{3}{2}}\alpha_{2i}^\dagger\alpha_{2i+1}\right)^{a^{i+1}}\\ &-& g \sum_{i=1}^{L}\left(-\omega^{-\frac{1}{2}}\alpha_{2i-1}^\dagger\alpha_{2i}\right)^{a^{-i}} - \    \mathcal{H}_{N,a}+ \mathrm{h.c.}
\end{eqnarray}
The nonlocal operator $\mathcal{H}_{N,a}$ arises from the boundary interaction in Eq. (\ref{4.13}):
\begin{equation}
    \mathcal{H}_{N,a} = \mathcal{Z}_{L}^{-a^{L+1}}\mathcal{Z}_1^a
    =
    c_{a,N}
    \left(\prod_{k=1}^{L} \mathcal{X}_k\right)^{a^{L+1}}
    (\alpha_{2L}^\dagger)^{a^{L+1}}\alpha_1^a,
\end{equation}
where $c_{a,N} = \omega^{\frac{a^{L+1}}{2}(N+2-a^{L+1})}$. When $a^L = 1 \bmod{N}$, this term simplifies to
\begin{equation}
    \mathcal{H}_{N,a} =
    \left[
    -\omega^{\frac{3}{2}}\alpha_{2L}^\dagger
    \left(\prod_{k=1}^{L} \mathcal{X}_k\right)\alpha_1
    \right]^a.
\end{equation}
Since $\prod_{k=1}^{L}\mathcal{X}_k = W^a$, the string operator can be incorporated into a boundary condition by defining
\begin{equation}
\label{4.16}
    \alpha_{2L+1} =
    \left(\prod_{k=1}^{L} \mathcal{X}_k\right)\alpha_1.
\end{equation}
The Hilbert space decomposes into $N$ symmetry sectors labeled by the eigenvalues of $W$. Because $a$ and $N$ are coprime, the eigenvalues of $\prod_{k=1}^{L}\mathcal{X}_k$ also run over $1,\omega,\ldots,\omega^{N-1}$. Within each symmetry sector, the string contributes a fixed phase to the boundary interaction, yielding a local parafermionic Hamiltonian with a boundary twist.

When $a^L \neq 1 \bmod{N}$, however, the string operator $\prod_{k=1}^{L}\mathcal{X}_k$ is no longer a symmetry, so this procedure of replacing it by a fixed eigenvalue does not apply.

\section{Partial parafermionic formulation for general $N$ and $a$ \label{secV}}

In this section, we return to the general case, in which $N$ and $a$ need not be coprime and may share prime factors. It is useful to define $N_a$ as the largest divisor of $N$ that is coprime to $a$. When $\gcd(a,N)=1$, we have $N_a=N$.

In Sec. \ref{IV}, we showed that the chain can be parafermionized using $\mathbb{Z}_N$ parafermions, with additional subtleties under PBC. Here, we show that a partial parafermionization is possible using $\mathbb{Z}_{N_a}$ parafermions. This construction elucidates the emergence of $\mathbb{Z}_{N_a}$ parafermion zero modes at the ends of an open chain at an exactly solvable point of the Hamiltonian.

In Sec. \ref{secVI}, we argue that this result is expected when the Hamiltonian in Eq. (\ref{3.1}) is interpreted as the boundary of a generalized toric code whose topological order depends on $a$ and $N$.

\subsection{Factorizing the Hilbert space}

First, we define $K$ by $N=N_aK$. By construction, $\gcd(N_a,a)=\gcd(N_a,K)=1$. This factorization allows us to identify the Hilbert space at each site with a tensor product:
\begin{equation}
    \mathbb{C}^N_i \simeq \mathbb{C}^{N_a}_i \otimes \mathbb{C}^{K}_i,
\end{equation}
where $i$ denotes the lattice site. Such decompositions have previously been used to express larger qudits in terms of smaller ones \cite{AVourdas_2003,AVourdas_2004}. Each clock basis state is identified with
\begin{equation}
    \ket{n_i \bmod N}_N
    \longleftrightarrow
    \ket{n_i \bmod N_a}_{N_a} \otimes \ket{n_i \bmod K}_K.
\end{equation}

In this basis, the generalized Pauli operators $X_i$ and $Z_i$ also decompose. The decomposition of $X_i$ is straightforward:
\begin{equation}
    X_i \longmapsto \tilde{X}_i \otimes \bar{X}_i.
\end{equation}
Its action on the basis states is
\begin{equation}
\begin{aligned}
    &(\tilde{X}_i \otimes \bar{X}_i)
    \left(\ket{n_i \bmod N_a}_{N_a} \otimes \ket{n_i \bmod K}_K\right)\\
    &\qquad =
    \ket{(n_i+1) \bmod N_a}_{N_a}
    \otimes
    \ket{(n_i+1) \bmod K}_K.
\end{aligned}
\end{equation}
The decomposition of $Z_i$, however, requires additional powers. Let $r=K^{-1}\bmod N_a$ and $s=N_a^{-1}\bmod K$ \footnote{More precisely, one can define a unitary operator $U$ such that $U\ket{n} =\ket{n}_{N_a}\otimes\ket{n}_K$. In such basis, the clock operators become $UXU^\dagger = \tilde{X}_i \otimes \bar{X}_i$ and $UZU^\dagger = \tilde{Z}_i^r \otimes \bar{Z}_i^s$.}. Then
\begin{equation}
    Z_i \longmapsto  \tilde{Z}_i^r \otimes \bar{Z}_i^s,
\end{equation}
where $\tilde{Z}_i$ and $\bar{Z}_i$ are $\mathbb{Z}_{N_a}$ and $\mathbb{Z}_K$ clock variables, respectively. The operators on the two factors satisfy
\begin{equation}
    \tilde{Z}_i\tilde{X}_i = \tilde{\omega}\tilde{X}_i\tilde{Z}_i,
    \qquad
    \tilde{\omega}=e^{\frac{2\pi i}{N_a}},
\end{equation}
\begin{equation}
    \bar{Z}_i\bar{X}_i = \bar{\omega}\bar{X}_i\bar{Z}_i,
    \qquad
    \bar{\omega}=e^{\frac{2\pi i}{K}}.
\end{equation}

These decompositions reproduce the original $\mathbb{Z}_N$ clock algebra:
\begin{equation}
\begin{aligned}
    &(\tilde{X}_i \otimes \bar{X}_i)
    (\tilde{Z}_i^r \otimes \bar{Z}_i^s)
    (\tilde{X}_i \otimes \bar{X}_i)^{-1}\\
    &\qquad =
    e^{-2\pi i\left(\frac{rK+sN_a}{N}\right)}
    (\tilde{Z}_i^r \otimes \bar{Z}_i^s)\\
    &\qquad =
    \omega^{-1}(\tilde{Z}_i^r \otimes \bar{Z}_i^s),
\end{aligned}
\end{equation}
since $rK+sN_a=1\bmod N$.

In this basis, the Hamiltonian in Eq. (\ref{3.1}), under OBC, becomes
\begin{equation}
\label{5.7}
    H = -\sum_{i=1}^{L-1}
    \tilde{Z}_i^{-ra}\tilde{Z}_{i+1}^{r}
    \otimes
    \bar{Z}_i^{-sa}\bar{Z}_{i+1}^{s}
    -g\sum_{i=1}^{L}\tilde{X}_i \otimes \bar{X}_i
    + \mathrm{h.c.}.
\end{equation}
Thus, we have decomposed each $\mathbb{Z}_N$ qudit into a $\mathbb{Z}_{N_a}$ qudit and a $\mathbb{Z}_K$ qudit. Our claim is that the $\mathbb{Z}_{N_a}$ factor can be parafermionized. This Hilbert-space factorization does not decouple the Hamiltonian: the two factors remain locally coupled. Nevertheless, we will show that the $K$ factor is topologically trivial.

\subsection{Topological and trivial sectors}

We first examine how the ground states decompose at $g=0$. The ground state condition in the $N_a$ factor is
\begin{equation}
    \tilde{Z}_i^{-ra}\tilde{Z}_{i+1}^{r}\ket{n_i,n_{i+1}}_{N_a}
    = \ket{n_i,n_{i+1}}_{N_a}.
\end{equation}
This gives
\begin{equation}
    e^{2\pi i\left(\frac{rn_{i+1}}{N_a}-\frac{ran_i}{N_a}\right)}=1.
\end{equation}
Since $r$ is coprime to $N_a$, this condition implies $n_{i+1}=an_i\bmod N_a$. Under OBC, it yields $N_a$ distinct ground states. Under PBC, however, the boundary interaction imposes the additional constraint $n_1=a^Ln_1\bmod N_a$. This congruence has $\gcd(a^L-1,N_a)$ solutions, giving the ground state degeneracy of this factor.

For the $K$ factor, the ground state condition is
\begin{equation}
    \bar{Z}_i^{-sa}\bar{Z}_{i+1}^{s}\ket{n_i,n_{i+1}}_K
    = \ket{n_i,n_{i+1}}_K.
\end{equation}
Similarly, we obtain $n_{i+1}=a^in_1\bmod K$. Since every prime factor of $K$ divides $a$, there exists a positive integer $p$ such that $a^p=0\bmod K$. Consequently, for a sufficiently long number of sites, each ground state of the $K$ factor can differ from zero only within a finite region:
\begin{equation}
    \ket{
    \underbrace{n_1,n_2,\ldots,n_p}_{\text{finite local region}},
    \underbrace{0,0,\ldots,0}_{\text{remaining sites}}
    }_K.
\end{equation}
Under OBC, this factor contributes $K$ ground states, yielding a total of $N=N_aK$ ground states. Under PBC, the $K$ factor has a unique ground state.

The nontrivial topological information is contained in the $N_a$ factor, whereas the ground state degeneracy of the $K$ factor is localized near the boundary.\footnote{More explicitly, the ground states of the $K$ factor are $\ket{t,ta,ta^2,\ldots,ta^{p-1},0,0,\ldots}_K$, where $t=0,1,\ldots,K-1$ and all entries are understood modulo $K$. The operator $V_K=\prod_{j=1}^{p}\bar{X}_j^{a^{j-1}}$, supported on the first $p$ sites, cyclically connects all $K$ degenerate ground states. Thus, this degeneracy is encoded locally.} Finally, the total ground state degeneracy under PBC is
\begin{equation}
    \mathrm{GSD}=\gcd(a^L-1,N_a)=\gcd(a^L-1,N),
\end{equation}
since $\gcd(a^L-1,K)=1$.

Under periodic boundary conditions, the symmetry operator in Eq. (\ref{3.2}) reduces to the operator
\begin{equation}
    U = \left(\prod_{i=1}^L \tilde{X}_i^{a^{i-1}}\right)^n.
\end{equation}
The $\mathbb{Z}_K$ sector of the symmetry operator trivializes due to the power of $n$.

\subsection{$\mathbb{Z}_{N_a}$ Parafermions}

In Sec. \ref{IV}, we saw that the coprimality of $a$ and $N$ is essential to the parafermionic construction there. Here, we apply this construction to the $\mathbb{Z}_{N_a}$ factor, since $\gcd(a,N_a)=1$.

Proceeding analogously to Eq. (\ref{4.1}), we introduce new clock variables,
\begin{equation}
    \mathcal{Z}_i = \tilde{Z}_i^{a^{-i}}, \qquad
    \mathcal{X}_i = \tilde{X}_i^{a^i},
\end{equation}
where $a^{-i}$ denotes the multiplicative inverse of $a^i$ modulo $N_a$. These variables satisfy the same clock algebra as the original operators. The Hamiltonian now reads
\begin{equation}
\begin{aligned}
    H ={}&
    -\sum_{i=1}^{L-1}
    \left(\mathcal{Z}_i^\dagger\mathcal{Z}_{i+1}\right)^{ra^{i+1}}
    \otimes \bar{Z}_i^{-sa}\bar{Z}_{i+1}^{s}\\
    &-\mathcal{Z}_{L}^{-ra^{L+1}}\mathcal{Z}_1^{ra}
    \otimes \bar{Z}_{L}^{-sa}\bar{Z}_{1}^{s}\\
    &-g\sum_{i=1}^{L}\mathcal{X}_i^{a^{-i}}\otimes\bar{X}_i
    +\mathrm{h.c.}
\end{aligned}
\end{equation}

We then map the $\mathbb{Z}_{N_a}$ clock operators to parafermions using the Fradkin-Kadanoff transformation:
\begin{equation}
    \beta_{2j-1} = \left(\prod_{k<j}\mathcal{X}_k\right)\mathcal{Z}_j,
    \qquad j\geq 1,
\end{equation}
\begin{equation}
    \beta_{2j} = \tilde{\omega}^{\frac{N_a+1}{2}}
    \left(\prod_{k<j}\mathcal{X}_k\right)\mathcal{Z}_j\mathcal{X}_j,
    \qquad j\geq 1.
\end{equation}
These parafermions satisfy the algebra
\begin{equation}
    \beta_j\beta_{j^\prime}
    =\tilde{\omega}^{\operatorname{sgn}(j^\prime-j)}
    \beta_{j^\prime}\beta_j.
\end{equation}

The Hamiltonian becomes
\begin{equation}
\label{5.18}
\begin{aligned}
    H ={}&
    -\sum_{i=1}^{L-1}
    \left(-\tilde{\omega}^{\frac{3}{2}}\beta_{2i}^\dagger\beta_{2i+1}\right)^{ra^{i+1}}
    \otimes\bar{Z}_i^{-sa}\bar{Z}_{i+1}^{s}
    -\mathcal{H}_{N,a}\\
    &-g\sum_{i=1}^{L}
    \left(-\tilde{\omega}^{-\frac{1}{2}}\beta_{2i-1}^\dagger\beta_{2i}\right)^{a^{-i}}
    \otimes\bar{X}_i
    +\mathrm{h.c.},
\end{aligned}
\end{equation}
where the nonlocal boundary term is
\begin{equation}
    \mathcal{H}_{N,a}=
    c_{a,N_a}
    \left(\prod_{i=1}^{L}\mathcal{X}_i\right)^{ra^{L+1}}
    (\beta_{2L}^\dagger)^{ra^{L+1}}\beta_1^{ra}
    \otimes\bar{Z}_{L}^{-sa}\bar{Z}_{1}^{s},
\end{equation}
with $c_{a,N_a}=\tilde{\omega}^{\frac{ra^{L+1}}{2}(N_a+2-ra^{L+1})}$.

When $\gcd(a^L-1,N_a)=N_a$, the operator $\prod_{i=1}^{L}\mathcal{X}_i$ is a symmetry of the Hamiltonian. Its eigenvalues label $N_a$ distinct symmetry sectors. Within each sector, the string operator contributes a fixed phase, and the boundary interaction becomes local. Labeling the resulting boundary twists by $k$, we can write
\begin{equation}
    \mathcal{H}_{N,a}=
    \tilde{\omega}^{k}
    \left(-\tilde{\omega}^{\frac{3}{2}}\beta_{2L}^\dagger\beta_1\right)^{ra}
    \otimes\bar{Z}_{L}^{-sa}\bar{Z}_{1}^{s},
\end{equation}
where $k=0,1,\ldots,N_a-1$. These twists correspond to the $N_a$ distinct boundary conditions of the parafermion chain.

Under open boundary conditions, the nonlocal boundary term is absent, and the Hamiltonian is fully local. In particular, at $g=0$,
\begin{equation}
    H = -\sum_{i=1}^{L-1}
    \left(-\tilde{\omega}^{\frac{3}{2}}\beta_{2i}^\dagger\beta_{2i+1}\right)^{ra^{i+1}}
    \otimes\bar{Z}_i^{-sa}\bar{Z}_{i+1}^{s}
    +\mathrm{h.c.}
\end{equation}

The two parafermions at the ends of the chain, $\beta_1$ and $\beta_{2L}$, commute with the Hamiltonian:
\begin{equation}
    [H,\beta_1]=[H,\beta_{2L}]=0, \qquad
    \beta_1^{N_a}=\beta_{2L}^{N_a}=1.
\end{equation}
They therefore constitute localized $\mathbb{Z}_{N_a}$ parafermion zero modes at the exactly solvable point $g=0$. As argued above, the $K$ factor does not support topological modes, so the nontrivial parafermionic end modes arise from the $N_a$ factor. In Appendix \ref{appendixA}, we present a concrete example with $N=14$ and $a=2$.

As stated at the beginning of this section, this construction applies to arbitrary values of $a$ and $N$: 
\begin{itemize}
\item[1.] When $a$ and $N$ are coprime, $K=1$, so the $\mathbb{Z}_K$ factor is trivial, and the construction reduces to that of Sec. \ref{IV};
\item[2.] Conversely, when $N_a=1$, every prime factor of $N$ also divides $a$, equivalently, $\text{rad}(N)$ divides $a$\footnote{The integer $\mathrm{rad}(N)$ is defined as the product of all distinct prime factors of $N$.}. In this case, the $\mathbb{Z}_{N_a}$ factor is trivial, and there are no nontrivial parafermion zero modes at the ends of the open chain. This is also the case where the symmetry becomes trivial, regardless of system size, and there is no phase transition.
\end{itemize}

\section{\label{secVI} The model as the boundary of a generalized toric code}

\begin{figure}
    \centering
    \includegraphics[width=1.0\linewidth]{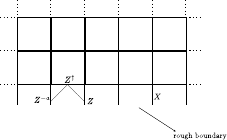}
    \caption{Rough boundary on the square lattice, and boundary operators illustrated by the bond of $Z$ and $X$ clock operators.}
    \label{boundary}
\end{figure}

In this section, we show that the generalized $\mathbb{Z}_N$ clock model can be understood as the boundary of a $(2+1)$-dimensional generalized $\mathbb{Z}_N$ toric code \cite{Watanabe:2021wwt}. This bulk model exhibits $\mathbb{Z}_{N_a}$ topological order, providing a bulk perspective on why the Fradkin-Kadanoff transformation involves $\mathbb{Z}_{N_a}$ parafermions.

The model is defined on a two-dimensional square lattice with qudits residing on its links. Its Hamiltonian is a sum of mutually commuting local stabilizers that generalize the plaquette and star operators of Kitaev’s toric code \cite{KITAEV20032}. These operators are given by
\begin{equation}
A_{\vec{s}}
=
\begin{tikzpicture}[baseline=-0.55ex, scale=0.9]

    \draw (-1.1,0) -- (1.4,0);
    \draw (0,-1.1) -- (0,1.1);

    \fill (0,0) circle (2pt);
    \node[below left=2pt] at (0,0) {$\vec{s}$};

    \node[fill=white,inner sep=2pt] at (-0.65,0) {$X$};
    \node[fill=white,inner sep=2pt] at (0,-0.65) {$X$};

    \node[fill=white,inner sep=2pt] at (0.80,0) {$X^{-a}$};
    \node[fill=white,inner sep=2pt] at (0.2,0.65) {$X^{-a}$};

\end{tikzpicture}, \quad \quad B_{\vec{p}}
=
\begin{tikzpicture}[baseline=(center.base), scale=0.9]

    \draw (0,0) rectangle (1.5,1.5);

    \node[inner sep=0pt] (center) at (0.75,0.75) {};

    \node[fill=white,inner sep=3pt] at (0.75,0) {$Z^{a}$};
    \node[fill=white,inner sep=3pt] at (0,0.75) {$Z^{-a}$};
    \node[fill=white,inner sep=3pt] at (0.75,1.5) {$Z^{\dagger}$};
    \node[fill=white,inner sep=3pt] at (1.5,0.75) {$Z$};

    \fill (0.75,0.75) circle (2pt);
    \node[below left=2pt] at (0.75,0.75) {$\vec {p}$};

\end{tikzpicture}
\end{equation}

The Hamiltonian is therefore given by $H=-\sum_{\vec{s}}A_{\vec{s}}-\sum_{\vec{p}}B_{\vec{p}} + \mathrm{h.c}$. We introduce a rough boundary along the bottom edge of the square lattice, as illustrated in Fig. \ref{boundary}. To construct the boundary Hamiltonian, we identify boundary operators that commute with the bulk stabilizers. For this rough boundary, these operators are given by
\begin{equation}
A^b_{\vec{s}} = 
\begin{tikzpicture}[baseline=(center.base), scale=0.9]

    \draw (0,0) -- (0,1.5);

    \node[fill=white,inner sep=3pt] at (0,0.75) {$X$};
\end{tikzpicture}, \quad \quad B^b_{p}=
\begin{tikzpicture}[baseline=(center.base), scale=0.9]

    \draw (0,0) -- (0,1.5);
    \draw (0,1.5) -- (1.5,1.5);
    \draw (1.5,1.5) -- (1.5,0);

    \node[inner sep=0pt] (center) at (0.75,0.75) {};

    \node[fill=white,inner sep=3pt] at (0,0.75) {$Z^{-a}$};
    \node[fill=white,inner sep=3pt] at (0.75,1.5) {$Z^{\dagger}$};
    \node[fill=white,inner sep=3pt] at (1.5,0.75) {$Z$};

\end{tikzpicture}
\end{equation}

These operators satisfy the same algebra as the corresponding terms in the Hamiltonian (\ref{3.1}). Moreover, upon projection onto the bulk ground state subspace, where the bulk stabilizers are fixed to their ground state values, they reproduce the spectrum of the generalized clock model within the boundary Hilbert space. The remaining degrees of freedom are the spins along the rough boundary, and the effective boundary Hamiltonian is given by
\begin{eqnarray}
\nonumber
    H &=& - \sum \begin{tikzpicture}[baseline=(center.base), scale=0.9]

    \draw (0,0) -- (0,1.5);

    \node[fill=white,inner sep=3pt] at (0,0.75) {$X$};
\end{tikzpicture} - \sum \begin{tikzpicture}[baseline=(center.base), scale=0.9]

    \draw (0,0) -- (0,1.5);
    \draw (1.0,1.5) -- (1.0,0);

    \node[inner sep=0pt] (center) at (0.75,0.75) {};

    \node[fill=white,inner sep=3pt] at (0,0.75) {$Z^{-a}$};
    \node[fill=white,inner sep=3pt] at (1.0,0.75) {$Z$};
\end{tikzpicture}\\ &=& -\sum_{i=1}^{L} X_i - \sum_{i=1}^{L} Z_i^{-a}Z_{i+1} +\mathrm{h.c.}.
\end{eqnarray}

Moreover, within the bulk ground state subspace, a weighted product of all bulk stabilizers imposes the constraint $U=1$ \cite{Seo:2024its}. Therefore, at low energies, the boundary of the generalized $\mathbb{Z}_N$ toric code is described by the generalized clock model restricted to the sector $U=1$.

\begin{figure*}[t]
    \centering
    \includegraphics[width=0.6\linewidth]{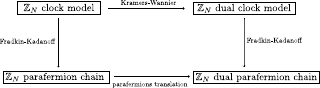}
    \caption{Diagram of maps between the $\mathbb{Z}_N$ generalized clock model, $\mathbb{Z}_N$ parafermion chain, and their duals.}
    \label{diagram}
\end{figure*}

For our purposes, the most relevant aspect of this perspective is the bulk topological order. Consider the generalized toric code on a torus with linear dimensions $L_1$ and $L_2$. Its ground state degeneracy is given by \cite{Watanabe:2021wwt}
\begin{equation}
\label{gsd}
    \text{GSD} = [\gcd(a^{L_1}-1,a^{L_2}-1,N_a)]^2.
\end{equation}

The maximum value of this degeneracy is $N_a^2$. The bulk realizes the quantum double $\mathcal{D}(\mathbb{Z}_{N_a})$ enriched by translation symmetry. Its anyons are generated by the electric and magnetic anyons $e$ and $m$, satisfying $e^{N_a}=m^{N_a}=1$.

The bulk topological order therefore suggests that, if a parafermionic formulation is possible, the relevant parafermions should be $\mathbb{Z}_{N_a}$ parafermions rather than $\mathbb{Z}_N$ parafermions, and under open boundary conditions, these are the emergent localized parafermions at the ends of the chain. For instance, the boundary of the usual $\mathbb{Z}_2$ toric code is effectively described by the transverse-field Ising model, which can be mapped to the Kitaev chain. Under open boundary conditions, the Kitaev chain can support a pair of Majorana zero modes, one localized at each end.

Another interesting case arises when every prime factor of $N$ also divides $a$. In this case, the degeneracy in Eq. (\ref{gsd}) equals one regardless of the torus dimensions, and the model exhibits no intrinsic topological order. This corresponds to $N_a=1$, for which no nontrivial parafermion zero modes occur at the ends of the generalized  clock model.

\section{From Kramers-Wannier duality to parafermion translation \label{secVII}}

In Ref. \cite{Seo:2024its}, it was shown that the generalized $N$-state clock model under periodic boundary conditions admits a non-invertible Kramers-Wannier duality. This duality maps the original clock variables to dual variables $\mu_i^x$ and $\mu_i^z$ through
\begin{equation}
\label{7.1}
    Z_i^aZ_{i+1}^{-1}\longmapsto\mu_{i+1}^x, \qquad
    X_i\longmapsto\mu_i^z(\mu_{i+1}^z)^{-a},
\end{equation}
where $\mu_i^z\mu_i^x=\omega\mu_i^x\mu_i^z$. The dual Hamiltonian is
\begin{equation}
\label{6.2}
    H = -\sum_{i=1}^{L}\mu_i^x
    -g\sum_{i=1}^{L}(\mu_i^z)^\dagger(\mu_{i+1}^z)^a
    +\mathrm{h.c.}
\end{equation}
A spatial reflection brings this Hamiltonian back to the form of the original Hamiltonian, with the bond and field couplings interchanged. Thus, recovering the original form requires mapping to the dual variables and then performing a spatial reflection.

In the transverse-field Ising model, Kramers-Wannier duality corresponds to a translation by one Majorana operator in the Kitaev chain. We will see that an analogous correspondence holds here, with some caveats.

To simplify the discussion, we consider the Hamiltonian in Eq. (\ref{4.13}) with $\gcd(a^L-1,N)=N$ and restrict to the symmetry sector $\prod_{i=1}^L X_i=1$, in which the duality becomes invertible \cite{Seo:2024its}. In this case, we consider the parafermionic Hamiltonian 
\begin{eqnarray}
\label{eq73}
\nonumber
    H = &-&\sum_{i=1}^{L}
    \left(-\omega^{\frac{3}{2}}\alpha_{2i}^\dagger\alpha_{2i+1}\right)^{a^{i+1}}+
    \\&-&g\sum_{i=1}^{L}
    \left(-\omega^{-\frac{1}{2}}\alpha_{2i-1}^\dagger\alpha_{2i}\right)^{a^{-i}}
    +\mathrm{h.c.}
\end{eqnarray}

\footnote{Note that, in the first sum, the last interaction contains $\alpha_{2L+1} = \prod_{k=1}^L X_k\alpha_1$. Before imposing $\prod_{k=1}^L X_k=1$, this product must move to the right, which leaves a phase factor.} The interpretation of duality as a translation, familiar from the Ising model, cannot be applied directly here. The main obstacle is that a naive translation of the parafermions under periodic boundary conditions is incompatible with their commutation relations, particularly across the boundary of the chain \cite{Yao:2020dqx}.

To interpret the duality in the parafermionic description, we first parafermionize the dual Hamiltonian in Eq. (\ref{6.2}). Following the same strategy as in Sec. \ref{IV}, we introduce a change of variables for the dual clock operators:
\begin{equation}
\label{7.4}
    \zeta_i^z = (\mu_i^z)^{a^{i}}, \quad \quad \zeta_i^x = (\mu_i^x)^{a^{-i}}.
\end{equation}
The sign of $i$ in the exponents differs from that in Eq. (\ref{4.1}), capturing the spatial reflection. The new variables $\zeta_i^x$ and $\zeta_i^z$ still satisfy the clock algebra and yield the Hamiltonian
\begin{equation}
    \tilde{H} = -\sum_{i=1}^{L}(\zeta_i^x)^{a^i}
    -g\sum_{i=1}^{L}\left[(\zeta_i^z)^\dagger\zeta_{i+1}^z\right]^{a^{-i}}
    +\mathrm{h.c.}
\end{equation}
We again apply the Fradkin-Kadanoff transformation, expressing the $\zeta_i^x$ and $\zeta_i^z$ variables in terms of parafermions $\tilde{\alpha}_i$. The Hamiltonian becomes
\begin{eqnarray}
\label{eq76}
\nonumber
    \tilde{H} = &-&\sum_{i=1}^{L}
    \left(-\omega^{-\frac{1}{2}}\tilde{\alpha}_{2i-1}^\dagger\tilde{\alpha}_{2i}\right)^{a^i}+
    \\&-&g\sum_{i=1}^{L}
    \left(-\omega^{\frac{3}{2}}\tilde{\alpha}_{2i}^\dagger\tilde{\alpha}_{2i+1}\right)^{a^{-i}}
    +\mathrm{h.c.}
\end{eqnarray}

We can now examine how Kramers-Wannier duality acts in the parafermionic description. In terms of the clock variables, the duality takes the form
\begin{equation}
     \mathcal{Z}_{i-1}\mathcal{Z}^\dagger_{i}\longmapsto \zeta_i^x, \quad \quad  \mathcal{X}_i \longmapsto \zeta_i^z(\zeta_{i+1}^z)^\dagger.
\end{equation}
These two relations lead to a parafermion translation of the bilinear operators
\begin{eqnarray}
\nonumber
   && \alpha^\dagger_j\alpha_{j+1} \longmapsto \tilde{\alpha}_{j+1}\tilde{\alpha}^\dagger_{j+2}, \quad \quad \text{if } j< 2L-1\\
    &&\alpha^\dagger_{2L-1}\alpha_{2L} \longmapsto \omega^{-1}\tilde{\alpha}_{2L}\tilde{\alpha}^\dagger_{1}.\\
    \nonumber
    && \alpha_{2L}^\dagger\alpha_1 \longmapsto \omega\, \tilde{\alpha}_1\tilde{\alpha}_2^\dagger
    \label{bilinear}
\end{eqnarray}

Thus, although a simple translation of the parafermions is incompatible with their algebra, we can still define a translation of parafermion bilinears.

This transformation maps the Hamiltonian $H$ in Eq. (\ref{eq73}) to the dual Hamiltonian $\tilde{H}$ in Eq. (\ref{eq76}), 
\begin{equation}
    H \longmapsto\tilde{H}.
\end{equation}

Kramers-Wannier duality is therefore represented by a parafermion translation between dual parafermion chains. Fig. \ref{diagram} shows a diagram of how the generalized clock model and the parafermion chain are mapped to each other under Kramers-Wannier duality and Fradkin-Kadanoff transformation.

\section{Conclusions}

In this work, we have investigated the parafermionic formulation of the generalized $N$-state clock model using the Fradkin-Kadanoff transformation. We have shown that a site dependent change of clock variables yields a local parafermionic Hamiltonian under open boundary conditions, with additional boundary subtleties under periodic boundary conditions. 

For general values of $N$ and $a$, this formulation involves identifying an $N_a$-dimensional factor in the Hilbert space at each site and applying the Fradkin-Kadanoff transformation to this factor using $\mathbb{Z}_{N_a}$ parafermions. At the exactly solvable point $g=0$, the resulting open chain supports parafermion zero modes of order $N_a$, one at each end, provided that $N_a>1$.

Additionally, we have shown that the generalized $N$-state clock model can be realized as the boundary of a generalized toric code whose bulk topological order is described by the quantum double $\mathcal{D}(\mathbb{Z}_{N_a})$. This connects the order of the parafermion zero modes to the bulk topological order.

Finally, we have shown that the Kramers-Wannier duality of the spin model corresponds to a translation of parafermion bilinears, mapping the original parafermion chain to a dual chain obtained from the dual spin model.

\vspace{0.3cm}

\begin{acknowledgments}

The author thanks Pedro Gomes and Guilherme Delfino for valuable discussions.
The author acknowledges the use of ChatGPT 5.6 Sol and 6 Astra for helpful discussions and assistance with the presentation. The research direction and main ideas were developed by the author, who takes full responsibility for the derivations, results, and final manuscript.

\end{acknowledgments}

\vspace{0.3cm}

\appendix

\section{\label{appendixA}Example: $N = 14$, $a=2$}

In this appendix, we construct the parafermionic formulation for the specific case $N=14$ and $a=2$. Here, $N_a=7$ and $K=2$, so $r=4$ and $s=1$. The Hamiltonian can therefore be written as
\begin{equation}
    H = -\sum_{i=1}^{L-1} \tilde{Z}^{-1}_i\tilde{Z}_{i+1}^{4} \otimes \bar{Z}_{i+1} -g\sum_{i=1}^{L}\tilde{X}_i \otimes \bar{X}_i + \mathrm{h.c.}.
\end{equation}
And the clock operators satisfy 
\begin{equation}
    \tilde{Z}_i\tilde{X}_i = e^{\frac{2\pi i}{7}}\tilde{X}_i \tilde{Z}_i, \quad \quad \bar{Z}_i\bar{X}_i = -\bar{X}_i \bar{Z}_i
\end{equation}

In this example, the ground state degeneracy of the $K$ factor is confined to the first site, consistent with its topologically trivial character. At $g=0$, its ground states are
\begin{equation}
    \ket{\psi_n}_K = \ket{n,0,\ldots,0}, \qquad
    \bar{Z}_i\ket{\psi_n}_K = \ket{\psi_n}_K
    \quad \text{for } i\geq 2,
\end{equation}
where $n=0,1,\ldots,K-1$.

We now define $\mathcal{Z}_i=\tilde{Z}_i^{2^{-i}}$ and $\mathcal{X}_i=\tilde{X}_i^{2^i}$, where $2^{-i}$ denotes the multiplicative inverse of $2^i$ modulo $7$. The Hamiltonian becomes
\begin{equation}
\begin{aligned}
    H ={}&
    -\sum_{i=1}^{L-1}
    \left(\mathcal{Z}_i^\dagger\mathcal{Z}_{i+1}\right)^{2^{i+3}}
    \otimes\bar{Z}_{i+1}+\\
    &-g\sum_{i=1}^{L}\mathcal{X}_i^{2^{-i}}\otimes\bar{X}_i
    +\mathrm{h.c.}
\end{aligned}
\end{equation}
These $\mathbb{Z}_7$ clock operators can be expressed in terms of $\mathbb{Z}_7$ parafermions $\beta_i$, yielding
\begin{equation}
\begin{aligned}
    H ={}&
    -\sum_{i=1}^{L-1}
    \left(-e^{\frac{3\pi i}{7}}\beta_{2i}^\dagger\beta_{2i+1}\right)^{2^{i+3}}
    \otimes\bar{Z}_{i+1}+\\
    &-g\sum_{i=1}^{L}
    \left(-e^{-\frac{\pi i}{7}}\beta_{2i-1}^\dagger\beta_{2i}\right)^{2^{-i}}
    \otimes\bar{X}_i
    +\mathrm{h.c.}
\end{aligned}
\end{equation}

At $g=0$, the open chain supports a pair of $\mathbb{Z}_7$ parafermion zero modes satisfying
\begin{equation}
    [H,\beta_1]=[H,\beta_{2L}]=0, \qquad
    \beta_1^7=\beta_{2L}^7=1.
\end{equation}
Thus, at the exactly solvable point $g=0$, a pair of zero modes is localized at the ends of the open chain.

\bibliographystyle{unsrt}
\bibliography{ref}

\end{document}